\documentclass[12pt]{article}
\usepackage{amsmath,amssymb,amsthm}
\usepackage{bbm}
\usepackage{graphicx}
\usepackage{epstopdf}
\usepackage{psfrag}

\newcommand{\Cx}{\mathbbm{C}}
\newcommand{\Ir}{\mathbbm{Z}}
\newcommand{\Nl}{\mathbbm{N}}

\newcommand{\Rl}{\mathbbm{R}}

\newcommand{\idty}{\mathbbm{1}}

 \DeclareMathOperator{\tr}{Tr}
\newcommand{\<}{\langle}
\renewcommand{\>}{\rangle}

\providecommand{\norm}[1]{\lVert#1\rVert}

\renewcommand{\b}[1]{\mathbf{#1}}
\newcommand{\bi}[1]{\boldsymbol{#1}}
\renewcommand{\c}[1]{\mathcal{#1}}

\newcommand{\s}[1]{\mathsf{#1}}
\renewcommand{\r}[1]{\mathrm{#1}}

\theoremstyle{definition}

\title{Connecting the von~Neumann and R\'enyi entropies for Fermions}
\date{}

\begin{document}
\maketitle
\begin{center}
Mark~Fannes\footnote{{\bf Email:} mark.fannes@fys.kuleuven.be}, Nicholas~Van~Ryn\footnote{{\bf Email:} nicholas.vanryn@fys.kuleuven.be}
\\[6pt]
Instituut voor Theoretische Fysica \\
KU Leuven, Belgium \\[6pt]
\end{center}
\setlength{\parindent}{0pt}
\setlength{\parskip}{6pt plus 2pt}
\begin{abstract}
\noindent We explore the relation between the von~Neumann entropy and the R\'enyi entropies of integer orders for shift-invariant quasi-free Fermionic lattice systems. We investigate approximating the von~Neumann entropy by a combination of integer-order R\'enyi entropies and give an estimate for the quality of such an approximation.\\

\noindent{\bf Keywords:} R\'enyi entropy, entropy, complete monotonicity, Fermionic systems, approximation scheme\\

\noindent{\bf PACS numbers:} 03.65.Db, 05.30.Ch, 05.70.Ce
\end{abstract}

\section{Introduction}
\label{s0}

The motivation for this paper is to better understand the relation between the average von~Neumann entropy and the average R\'enyi entropies of integer order. Entropies are non-local characteristics of a state and are, e.g.,\ an essential input in the variational principle for thermal equilibrium. Restricting the variational principle to specific classes of states leads to well-known approximations like mean-field or Hartree-Fock~\cite{martin1990problemes}. More refined approximation schemes like matrix product states for computing ground states of quantum spin chains turned out to be quite effective~\cite{PhysRevLett.107.070601,PhysRevLett.75.3537}. Extensions to higher dimensional quantum spin lattices are currently investigated~\cite{springerlink:10.1007/s10955-011-0237-4} and one might wonder about using general finitely correlated states for thermal states. Computing the mean von~Neumann entropy for a general, shift-invariant, finitely correlated state is, however, still an open problem~\cite{fannes1992}.

In~\cite{Renyi:1961ty}, R\'enyi introduced a generalised entropy of order $\alpha$ defined as
\begin{equation}
\s S_\rho(\alpha) := - \frac{1}{\alpha-1}\, \tr \rho^\alpha,
\end{equation}
where $\rho$ is a density matrix, $\alpha \not= 1$, and $\alpha\ge 0$. The von~Neumann entropy is obtained as the limit
\begin{equation}
\s S_\rho = \lim_{\alpha \to 1} \s S_\rho(\alpha) = - \tr \rho \log\rho.
\end{equation}
R\'enyi entropies of integer order $\alpha \in \{2,3,\ldots\}$ can often be easily computed using several copies of the system, by way of the so-called replica trick:
\begin{equation}
\tr \rho^n = \tr \underbrace{\rho \otimes \rho \otimes \cdots \otimes \rho}_{n\ \text{times}}\, T_n
\end{equation}
where $T_n$ is the cyclic shift on the $n$-fold tensor power of the original system. For an example of this replica trick applied to a spin glass, see~\cite{0305-4470-11-5-028}. The question is then how to reconstruct the von~Neumann entropy given the R\'enyi entropies of integer order $2,3,\ldots$

In fact, the relevant quantity for shift-invariant states is the average R\'enyi entropy. Unfortunately, these densities pose several serious problems with regard to existence and continuity. In general, they simply don't exist and they are also not affine on convex subsets of shift-invariant states. Their use is therefore limited to states with strong clustering. Moreover, it is completely unclear whether the knowledge of integer-order average R\'enyi entropies uniquely determines the average von~Neumann entropy.

A number of papers have considered relations between the integer-order R\'enyi and von~Neumann entropies~\cite{Harremoes2009f,PhysRevA.70.022316,springerlink:10.1023/A:1025128024427}. These relations don't always scale properly with the system size and therefore don't necessarily survive on the level of densities. There is certainly no general procedure, even under strong assumptions on clustering, for passing from integer-order R\'enyi to von~Neumann entropies. Further suggested reading on the subject can be found in \cite{bengtsson, Ohya1993, Petz}.

In this paper we consider the case of shift-invariant quasi-free Fermionic states on a lattice. We show that such a reconstruction procedure exists in this case and obtain some simple approximations in terms of the first few R\'enyi densities.

The paper is organised as follows: in Section~\ref{s1}, we remind the reader of the description of Fermions on a lattice and introduce the notation. Section~\ref{s2} gives the expression for the average R\'enyi entropies of quasi-free states. In Section~\ref{s3}, we introduce a completely monotonic entropy function which is then used in Section~\ref{s4} to reconstruct the von~Neumann entropy. Finally, we provide an explicit approximation scheme in Section~\ref{s5}

\section{Fermions on a lattice}
\label{s1}

We consider a system of Fermions living on some Bravais lattice $\c L$ in $\Rl^d$:
\begin{equation}
\c L = \{n_1 \varepsilon_1 + n_2 \varepsilon_2 + \cdots + n_d \varepsilon_d \mid n_1,n_2,\ldots, n_d \in \Ir \} = \{\b n \cdot \bi\varepsilon \mid \b n \in \Ir^d\},
\end{equation}
where $\{\varepsilon_j\}$ are the primitive vectors of the lattice. For our purposes, we can identify $\b n \cdot \bi\varepsilon$ with $\b n$.

Fermions on the lattice are described by smeared out creation and annihilation operators $c^\dagger$ and $c$ obeying the canonical anticommutation relations. The smearing is by square summable sequences on the lattice:
\begin{equation}
\ell^2(\Ir^d) \ni \varphi \mapsto c^\dagger(\varphi)\enskip \text{is $\Cx$-linear}
\end{equation}
and the anticommutation relations are
\begin{equation}
\{c(\varphi), c(\psi)\} = 0 \enskip\text{and}\enskip \{c(\varphi),c^\dagger(\psi)\} = \<\varphi,\psi\> \idty.
\end{equation}
The creation and annihilation operators generate the C*-algebra $\c A(\Ir^d)$ of canonical anticommutation relations (CAR) on $\Ir^d$.

Lattice translations induce $*$-automorphisms on $\c A(\Ir^d)$ by extending
\begin{equation}
c(\varphi) \mapsto c(U(\b n)\varphi).
\end{equation}
Here $U(\b n)$ is the unitary shift on $\ell^2(\Ir^d)$ by $\b n \in \Ir^d$
\begin{equation}
\bigl( U(\b n)\, \varphi \bigr)(\b k) = \varphi( \b k - \b n).
\label{1.1}
\end{equation}

We also need quasi-free states $\omega_Q$, sometimes called Gaussian Fermionic states. They are uniquely determined by their two-point correlation functions
\begin{equation}
(\varphi,\psi) \mapsto \omega\bigl(c^\dagger(\varphi) c(\psi) \bigr),\enskip \varphi,\psi \in \ell^2(\Ir^d).
\end{equation}
All expectations of monomials in creation and annihilation operators vanish except for
\begin{equation}
\omega\bigl( c^\dagger(\varphi_1) \cdots c^\dagger(\varphi_n) c(\psi_n) \cdots c(\psi_1) \bigr) = \det \Bigl( \bigl[ \omega\bigl(c^\dagger(\varphi_k) c(\psi_\ell) \bigr) \bigr]_{k\ell} \Bigr).
\label{1.2}
\end{equation}
Due to the complex linearity of $c^\dagger$, and the conjugate linearity of $c$,
\begin{equation}
(\varphi,\psi) \mapsto \omega\bigl(c^\dagger(\varphi) c(\psi) \bigr)
\label{1.3}
\end{equation}
is a sesquilinear form on $\ell^2(\Ir^d)$. One can show that the necessary and sufficient condition for~(\ref{1.2}) to define a state is that the form~(\ref{1.3}) corresponds to a linear operator $Q$ on $\ell^2(\Ir^d)$ such that $0 \le Q \le \idty$:
\begin{equation}
\omega\bigl( c^\dagger(\varphi) c(\psi) \bigr) = \<\psi, Q\, \varphi \>.
\end{equation}
The operator $Q$, called the symbol, defines the state $\omega_Q$, where the notation stresses the dependence of the state $\omega$ on $Q$.

A quasi-free state $\omega_Q$ is shift-invariant if and only if $Q$ commutes with the lattice-shift unitaries $U(\b n)$ defined in~(\ref{1.1}). In the standard basis $\{e_{\b j}\}$ of $\ell^2(\Ir^d)$ this is equivalent with
\begin{equation}
\<e_{\b j}, Q\, e_{\b k} \> = \<e_{\b j + \b n}, Q\, e_{\b k + \b n} \>,\enskip \b j, \b k, \b n \in \Ir^d.
\end{equation}
If $F$ denotes the unitary Fourier transformation
\begin{equation}
(F\,\varphi)(\b x) := \sum_{\b n \in \Ir^d} \varphi(\b n)\, \r e^{2\pi i \b n \cdot \b x},\enskip \b x \in [0,1]^d
\end{equation}
this is equivalent to
\begin{equation}
F\, Q = q\, F
\label{1.4}
\end{equation}
where $q$ is the multiplication operator with the function
\begin{equation}
q(\b x) = \sum_{\b j \in \Ir^d} \<e_{\b 0}, Q\, e_{\b j} \>\, \r e^{2\pi i \b j \cdot \b x}
\label{1.5}
\end{equation}
on $\c L^2([0,1]^d)$. As $0 \le Q \le \idty$, $q$ takes values in $[0,1]$. For more details on these matters we refer to~\cite{dierckx}.

\section{R\'enyi entropies}
\label{s2}

Let $\Lambda$ be a finite subset of $\Ir^d$. The local algebra $\c A(\Lambda)$ is generated by the creation and annihilation operators with smearing functions supported in $\Lambda$. This algebra is isomorphic to the algebra of matrices of dimension $2^{\#(\Lambda)}$ and has therefore, up to unitary equivalence, a unique irreducible representation. This allows one to assign to any state $\omega$ on $\c A(\Lambda)$ the R\'enyi entropies
\begin{equation}
\s S_\omega(\alpha) := - \frac{1}{\alpha - 1}\, \log\tr \rho^\alpha,\enskip \alpha > 0.
\label{2.1}
\end{equation}
Here $\rho$ is the density matrix defining $\omega$ in an irreducible representation of $\c A(\Lambda)$.
For $\alpha = 1$, the expression in~(\ref{2.1}) has to be replaced by its limit, the von~Neumann entropy
\begin{equation}
\s S_\omega := \s S_\omega(1) = - \tr (\rho \log\rho).
\end{equation}

The local R\'enyi entropies of quasi-free states can be readily expressed in terms of the local restrictions of their corresponding symbols $Q$. Moreover, one can show that for shift-invariant quasi-free states the limiting average R\'enyi entropies in the sense of growing boxes exist and one obtains an explicit expression in terms of the Fourier transform $q$:
\begin{equation}
\s s_q(\alpha) = - \frac{1}{\alpha-1}\, \int_{[0,1]^d} \!d\b x\, \log\bigl( q^\alpha + (1-q)^\alpha \bigr).
\label{2.2}
\end{equation}
In the case of the von~Neumann entropy density, this expression must again be understood as
\begin{equation}
\s s_q = - \int_{[0,1]^d} \!d\b x\, \bigl[ q\log q  + (1-q)\log(1-q) \bigr].
\end{equation}

\section{A completely monotonic entropy}
\label{s3}

It is known that for a general density matrix $\rho$ in a matrix algebra
\begin{equation}
\alpha \in ]0,\infty[ \mapsto - \frac{1}{\alpha - 1}\, \log\tr \rho^\alpha
\end{equation}
is a monotonically decreasing function, in particular
\begin{equation}
\s S_\rho \ge \s S_\rho(2) \ge \s S_\rho(3) \ge \cdots
\end{equation}
This ordering then extends to densities, provided they exist. This is certainly the case for shift-invariant quasi-free states, where it is not hard to compute the asymptotic value
\begin{equation}
\s s_q(\infty) = \lim_{\alpha \to \infty} \s s_q(\alpha) = - \int_{[0,1]^d} \!d\b x\, \log\bigl[\max\{ q, (1-q)\} \bigr].
\label{2.3}
\end{equation}

We now introduce a modified entropy-like function:
\begin{equation}
g_q(\alpha) := \s s_q(\infty) - \tfrac{\alpha-1}{\alpha}\, \s s_q(\alpha),\enskip \alpha > 0.
\label{3.1}
\end{equation}
To avoid technical complications, we assume from now on that $q=0$ or $q=1$ on a set of measure zero. This is, e.g.,\ true for thermal states of quadratic interactions at non-zero temperature. The general case can be handled by removing the union of the kernels of $q$ and $1-q$ from $[0,1]^d$. One now easily verifies that
\begin{equation}
g_q(\alpha) = \frac{1}{\alpha}\, \int_{[0,1]^d} \!d\b x\, \log[ 1 + \exp(-\alpha h)],
\label{3.2}
\end{equation}
where we have introduced the function
\begin{equation}
h: [0,1]^d \to \Rl^+: h := - \log\Bigl(\min\Bigl\{ \frac{q}{1-q}, \frac{1-q}{q} \Bigr\} \Bigr).
\end{equation}

A function $f:]0,\infty[ \to \Rl$ is called completely monotonic if its derivatives to all orders exist and if
\begin{equation}
(-1)^n f^{(n)} \ge 0,\enskip n=0,1,2,\ldots
\end{equation}
Bernstein's theorem \cite{widder2010laplace} characterises completely monotonic functions as the Laplace transforms of positive measures that don't grow too fast at infinity. We show here that, for a given $q$,
\begin{equation}
\alpha \in ]0,\infty[ \mapsto g_q(\alpha)
\end{equation}
is completely monotonic by computing its inverse Laplace transform.

\begin{figure}[t]
\input{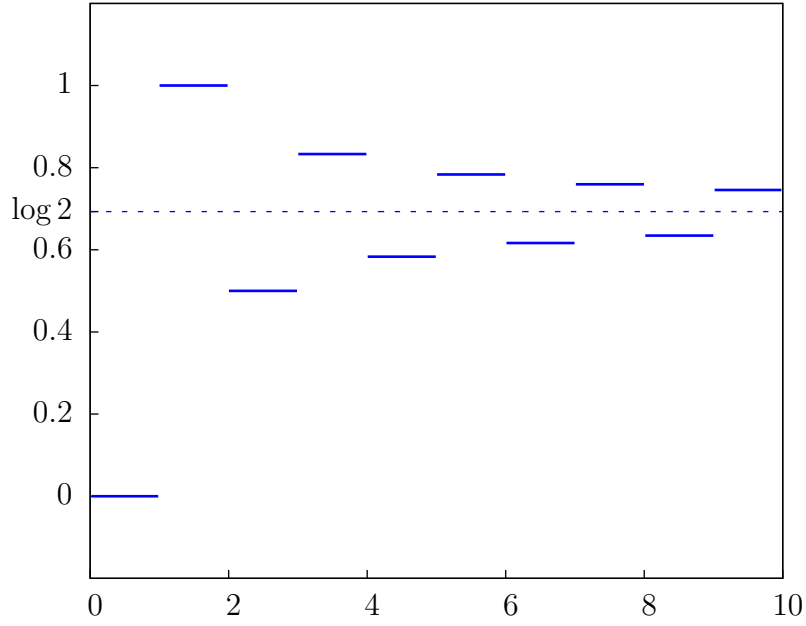}
\caption{The inverse Laplace transform of $s \mapsto \log(1 + \exp(-s))/s$~\label{fig1}}
\end{figure}

First, consider the function $k$ on $\Rl^+$:
\begin{equation}
k(t) = \sum_{\substack{j \in \Nl_0 \\ j \le t}} \frac{(-1)^{j+1}}{j},\enskip t \ge 0.
\end{equation}
This function is piecewise constant, non-negative, continuous from the right, and tends to $\log 2$, see Fig.~\ref{fig1}. As $k$ is of exponential order 0, we may compute its Laplace transform for all $s \in \Cx$ with $\Re(s) > 0$ by evaluating the integral
\begin{align}
\c L(k)(s) 
&= \int_{\Rl^+} \!dt\, k(t)\, \r e^{-st} = \sum_{\ell=1}^\infty \sum_{j=1}^\ell \frac{(-1)^{j+1}}{j} \int_\ell^{\ell+1} \!dt\, \r e^{-st} \nonumber \\
&= \sum_{\ell=1}^\infty \frac{1}{s}\, \Bigl( \r e^{-\ell s} - \r e^{-(\ell+1)s} \Bigr) \Bigl(\sum_{j=1}^\ell \frac{(-1)^{j+1}}{j} \Bigr) \nonumber\\ 
&= \frac{1}{s}\, \sum_{\ell=1}^\infty \frac{(-1)^{\ell+1}}{\ell}\, \r e^{-\ell s} = \frac{1}{s}\, \log\bigl( 1 + \exp(-s) \bigr).
\end{align}
This shows that
\begin{equation}
s >0 \mapsto \frac{1}{s}\, \log\bigl( 1 + \exp(-s) \bigr)
\end{equation}
is completely monotonic.

As complete monotonicity is preserved by rescaling the argument, rescaling the function, and addition, we conclude from~(\ref{3.2}) that
\begin{equation}
\alpha \in ]0,\infty[ \mapsto g_q(\alpha)
\end{equation}
is completely monotonic. We have, in fact, a rather explicit expression for the inverse Laplace transform of $g_q$:
\begin{equation}
\c L^{-1}\bigl( g_q \bigr)(t) = \int_{[0,1]^d} \!d\b x\, k\Bigl( \frac{t}{h} \Bigr).
\end{equation}

\section{Reconstructing the von~Neumann entropy}
\label{s4}

In general, one cannot hope to uniquely reconstruct a function $g$ that is analytic in $\{z \in \Cx \mid \Re(z) > 0 \}$ given the values of $g$ on $\Nl_0$. This is nevertheless often attempted in statistical mechanics of disordered systems where one tries to reconstruct the von~Neumann entropy given the R\'enyi entropies of order 2, 3, \dots which can be computed using the replica trick. Suppose, however, that there exists a non-negative measurable function $G$ of exponential order 0 on $\Rl^+$ such that
\begin{equation}
g(\alpha) = \int_0^\infty \!dt\, G(t)\, \r e^{-\alpha t},\enskip \alpha > 0
\end{equation}
where $g$ is related to the entropy densities $\s s$ as in~(\ref{3.1}). We then have
\begin{equation}
\s s(\alpha) = \frac{\alpha}{\alpha-1}\, \int_0^\infty \!dt\, G(t)\, \Bigl( \r e^{-t} - \r e^{-\alpha t} \Bigr),\enskip \alpha \ne 1
\label{4.1}
\end{equation}
and
\begin{equation}
\s s = \s s(1) = \int_0^\infty \!dt\, G(t)\, t\, \r e^{-t}.
\label{4.2}
\end{equation}

Suppose that we know the integer-order entropies $\s s(n)$ for $n = 2,3,\ldots$ and that we wish to reconstruct the von~Neumann entropy $\s s$ given~(\ref{4.1}) and~(\ref{4.2}). The vector space generated by the functions
\begin{equation}
f_n: t \in \Rl^+ \mapsto \frac{n}{n-1}\, \bigl( \r e^{-t} - \r e^{-nt} \bigr),\enskip n = 2,3,\ldots
\label{4.3}
\end{equation}
is actually an algebra of continuous functions that vanish at 0 and at $\infty$. It is closed under complex conjugation and it separates the points in $\Rl^+$. We can therefore approximate
\begin{equation}
t \in \Rl^+ \mapsto t\, \r e^{-t}
\end{equation}
uniformly to arbitrary precision by a linear combination of the $f_n$ using the Stone-Weierstrass theorem.

Given a $d$-dimensional density matrix $\rho$ and $d-1$ R\'enyi entropies of integer order in $\{2,3,\ldots\}$ we can uniquely reconstruct the ordered eigenvalues of $\rho$. This means that two density matrices with the same integer R\'enyi entropies are related by a unitary transformation.

The mean R\'enyi entropies of a shift-invariant quasi-free state are of the form $\int_{[0,1]^d} \!d\b x\, f(q)$ where $f$ is a bounded measurable function on $[0,1]$. We can associate a distribution function to a symbol $q$ in Fourier space as follows,
\begin{equation}
\gamma_q: y \in [0,1] \mapsto \int_{\substack{[0,1]^d \\ q(\b x) \le y}} \!d\b x.
\label{4.4}
\end{equation}
We can then write
\begin{equation}
\int_{[0,1]^d} \!d\b x\, f(q) = \int_0^1 \!d\gamma_q(y)\, f(y).
\end{equation}
Obviously, different symbols may yield the same distribution function and therefore the same mean R\'enyi entropies. Knowledge of the integer mean entropies with $\alpha \in \{2,3,\ldots\}$ actually determines the distribution function except for contributions at 0 or 1 which don't effectively contribute to any mean entropy for $\alpha > 0$.

Suppose that $t$ is a Lebesgue measure preserving rearrangement of $[0,1]^d$, i.e.\ a transformation of $[0,1]^d$ such that
\begin{equation}
\int_{\b x \in \Lambda} \!d\b x = \int_{t(\b x) \in \Lambda} \!d\b x
\end{equation}
for every measurable set $\Lambda$. The mapping
\begin{equation}
U_t: \varphi \mapsto \varphi \circ t
\end{equation}
is a unitary transformation of $\c L^2([0,1]^d)$ such that for a multiplication operator by $q$
\begin{equation}
U_t\, q = (q \circ t)\, U_t.
\end{equation}
The Fourier transformed symbols $q$ and $q \circ t$ define different shift-invariant quasi-free states on $\c A(\Ir^d)$ which are related by the automorphism defined through $U_t$.\\

The distribution functions~(\ref{4.4}) of these states coincide and therefore have the same mean entropies. This situation can be seen as a partial quasi-free analogue of the density matrix case discussed above. There are, however, plenty of  symbols that yield the same mean entropies and that are not related by a rearrangement.

\section{Explicit approximations}
\label{s5}

Let us extend the notation in (\ref{4.3}) by defining
\begin{equation}
f_1(t) = t \r e^{-t}.
\end{equation}
A simple-minded $n$-term approximation of $f_1$ consists in finding the minimisers of
\begin{equation}
(\gamma_1, \ldots, \gamma_n) \mapsto \norm{f_1 - \gamma_1\, f_2 -\cdots - \gamma_n\, f_{n+1}}_\infty
\label{4.5}
\end{equation}
by demanding that all extrema of the above function are equal in magnitude.
The first few optimal approximations are
\begin{equation}
\begin{split}
&\norm{f_1 - 0.800\, f_2}_\infty = 0.09 \\
&\norm{f_1 - 2.219\, f_2 + 1.314\, f_3}_\infty = 0.04 \\
&\norm{f_1 - 4.233\, f_2 + 6.133\, f_3 - 2.850\, f_4}_\infty = 0.02\\
&\norm{f_1 - 6.833\, f_2 + 17.498\, f_3 - 18.780\, f_4 + 7.148\, f_5}_\infty = 0.01.
\end{split}
\label{5.1}
\end{equation}
In comparison, $\norm{f_1}_\infty \approx 0.368$.

\begin{figure}[t]
    \input{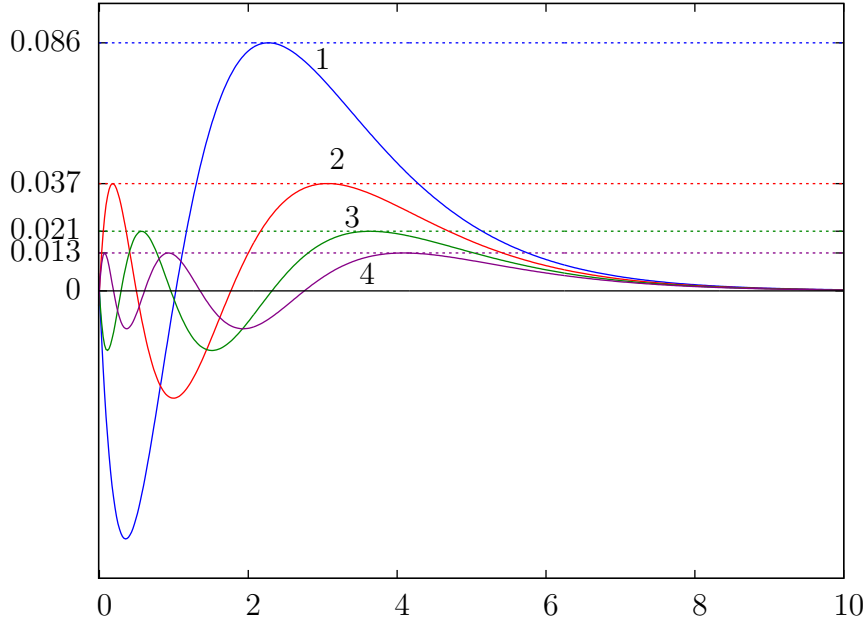}
    \caption{Difference between $t\r e^{-t}$ and its $1$, $2$, $3$ and $4$-term approximations as indicated, given by Equation (\ref{4.5}) \label{fig2}}
\end{figure}

This corresponds to the approximations of the mean von~Neumann entropy by the first few mean R\'enyi entropies
\begin{equation}
\begin{split}
\s s_q
&\approx 0.800\, \s s_q(2)\\
&\approx 2.219\, \s s_q(2) - 1.314\, \s s_q(3) \\
&\approx 4.233\, \s s_q(2) - 6.133\, \s s_q(3) + 2.850\, \s s_q(4)\\
&\approx 6.833\, \s s_q(2) - 17.498\, \s s_q(3) + 18.780\, \s s_q(4) - 7.148\, \s s_q(5)
\end{split}
\label{5.2}
\end{equation}
The single term approximation is certainly terrible because we know from the monotonicity of the R\'enyi densities that
\begin{equation}
\s s_q \ge \s s_q(2).
\end{equation}
The multi-term approximations are, however, no longer direct consequences of the monotonicity of the R\'enyi densities. The quality of~(\ref{5.2}) is not easily established because the measure $G(t) dt$ is generally unbounded.

Under our assumption on $q$, the function $G$ in~(\ref{4.1}) tends to $\log 2$ at infinity. As the supremum norm is attained close to the origin, we cannot expect to obtain a very good approximation this way. A better approach, especially at high temperatures, is to subtract the asymptotic value $\log 2$ from $G$ and then use~(\ref{5.1}). This yields
\begin{equation}
\begin{split}
\s s_q
&\approx 0.200\, \log 2 + 0.800\, \s s_q(2) \\
&\approx 0.095\, \log 2 + 2.192\, \s s_q(2) - 1.314\, \s s_q(3) \\
&\approx 0.050\, \log 2 + 4.233\, \s s_q(2) - 6.133\, \s s_q(3) + 2.850\, \s s_q(4)\\
&\approx 0.033\, \log 2 + 6.833\, \s s_q(2) - 17.498\, \s s_q(3) + 18.785\, \s s_q(4) - 7.148\, \s s_q(5)
\end{split}
\end{equation}
One can, however, still not hope that $G - \log 2$ is integrable.

A controllable approximation scheme can be set up by using R\'enyi densities of order less than one and using the uniform bound
\begin{equation}
\s s_q(\alpha) = \frac{\alpha}{\alpha - 1}\, \int_0^\infty \!dt\, G(t)\, \Bigl( \r e^{-t} - \r e^{-\alpha t} \Bigr) \le \log 2.
\end{equation}
Extending the definition of $f_n$ in~(\ref{4.3}) to general $\alpha \in \Rl^+$ we write
\begin{align}
&\Bigl| \s s_q - \gamma_1\, \s s_q(2) -\cdots - \gamma_n\, \s s_q(n+1) \Bigr| \nonumber\\
&\quad= \Bigl| \int_0^\infty \!dt\, G(t)\, \Bigl( f_1(t) - \gamma_1\, f_2(t) -\cdots - \gamma_n\, f_{n+1}(t) \Bigr) \Bigr| \nonumber \\
&\quad= \Bigl| \int_0^\infty \!dt\, G(t)\, f_\alpha(t)\, \Bigl( \frac{f_1(t)}{f_\alpha(t)} - \gamma_1\, \frac{f_2(t)}{f_\alpha(t)} -\cdots - \gamma_n\, \frac{f_{n+1}(t)}{f_\alpha(t)} \Bigr) \Bigr| \nonumber \\
&\quad\le \Bigl\Vert \frac{f_1}{f_\alpha} - \gamma_1\, \frac{f_2}{f_\alpha} -\cdots - \gamma_n\, \frac{f_{n+1}}{f_\alpha}\Bigr\Vert_\infty\, \log 2.
\label{5.3}
\end{align}

It now remains to minimise the bound~(\ref{5.3}) for a given $n$ with respect to $\gamma_1, \ldots, \gamma_n$ and $\alpha \in ]0,1[$. This leads to
\begin{equation}
\begin{split}
&\bigl| \s s_q - 0.666\, \s s_q(2) \bigr| \le 0.35 \\
&\bigl| \s s_q - 1.938\, \s s_q(2) +1.005\, \s s_q(3) \bigr| \le 0.19 \\
&\bigl| \s s_q - 3.892\, \s s_q(2) + 4.967\, \s s_q(3) - 2.048\, \s s_q(4) \bigr| \le 0.12 \\
&\bigl| \s s_q - 6.556\, \s s_q(2) + 15.064\, \s s_q(3) - 14.413\, \s s_q(4) + 4.923\, \s s_q(5) \bigr| \le 0.08 \\
&\cdots \\
&\bigl| \s s_q - 37.181\, \s s_q(2) + 529.415\, \s s_q(3) - 3\,846.261\, \s s_q(4) + 16\,301.725\, \s s_q(5) \\
&\phantom{\bigl| \s s_q}- 43\,168.833\, \s s_q(6) + 73\,647.855\, \s s_q(7) - 80\,999.681\, \s s_q(8) + 55\,517.489\, \s s_q(9) \\
&\phantom{\bigl| \s s_q}- 21\,580.373\, \s s_q(10) + 3\,634.848\, \s s_q(11) \bigr| \le 0.03
\end{split}
\label{5.4}
\end{equation}

\begin{figure}[h]
    \input{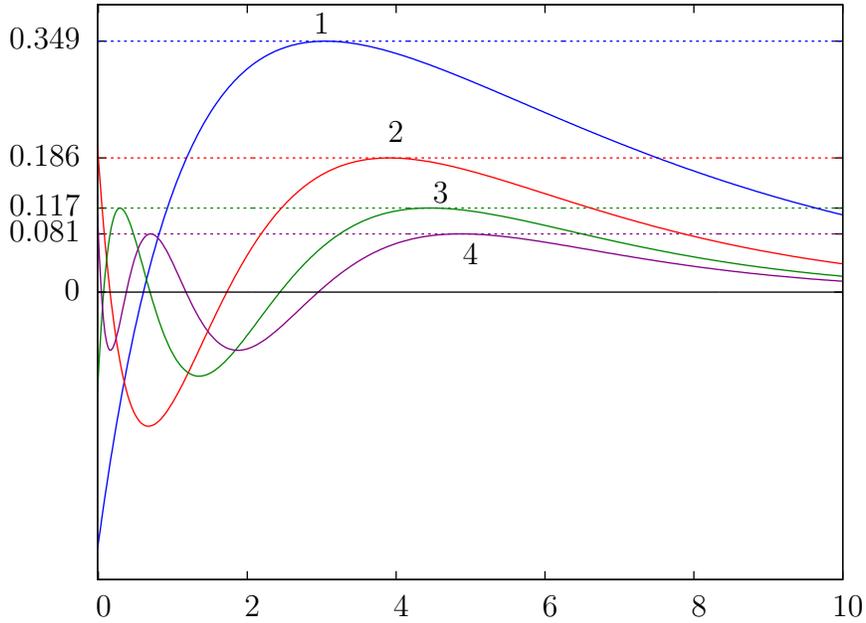}
    \caption{Difference between $f_1/f_\alpha$ and its controlled $1$, $2$, $3$ and $4$-term approximations re-scaled by $f_\alpha$, given by Equations (\ref{5.3}) \label{fig3}}
\end{figure}

The corresponding values for $\alpha$ are: 0.661, 0.515, 0.435, 0.384, and 0.261 respectively.

\section{Conclusion}

In this paper, we consider the reconstruction of the average von~Neumann entropy in terms of average R\'enyi entropies of integer order, which, for some classes of states, are computed rather easily. Obtaining general estimates and relations for these quantities is a long-standing problem.

We have restricted our attention to shift-invariant quasi-free Fermionic states on a regular lattice and obtain two results. Firstly, we proved that the mean von~Neumann entropy is reconstructible in terms of mean R\'enyi entropies of integer order. Next, we set up an explicit approximation scheme. This scheme, including the controlled approximations~(\ref{5.4}), relies only on the validity of the representation~(\ref{4.1}) and is therefore applicable to general systems for which~(\ref{4.1}) holds and is not only limited to the shift-invariant quasi-free states considered here. Of course, the $\log 2$ asymptotic limit given in the controlled approximation scheme should be adjusted according to the dimensionality of the states in question. It should be noted that the coefficients in~(\ref{5.1}) and~(\ref{5.4}) are not easily expressible in a general analytical form. Moreover, the obtained approximations are generally neither upper nor lower bounds.

\noindent
\textbf{Acknowledgements}
This work was partially funded by the Belgian Interuniversity Attraction Poles Program
P6/02 and by the FWO Vlaanderen project G040710N.

\bibliographystyle{plain}
\bibliography{references}

\end{document}